\documentclass[twocolumn,showpacs,preprintnumbers,prb,fleqn,eps]{revtex4}
\usepackage{epsfig}
\usepackage{graphicx}
\usepackage{amsfonts}
\begin{document}

\title{Thermal entanglement witness for materials with variable local spin lengths}

\author{Andre M. C. Souza$^{1}$ and Francisco A. G. Almeida$^{2}$}
\affiliation{$^{1}$Departamento de Fisica, Universidade Federal de
Sergipe, 49100-000 Sao Cristovao-SE, Brazil}
\affiliation{$^{2}$Departamento de Fisica, Universidade Federal de
Pernambuco, 50670-901 Recife-PE, Brazil}

\date{\today}

\begin{abstract}

We show that the thermal entanglement in a spin system using only
magnetic susceptibility measurements is restricted to the insulator
materials. We develop a generalization of the thermal entanglement
witness that allows us to get information about the system
entanglement with variable local spin lengths that can be used
experimentally in conductor or insulator materials. As an
application, we study thermal entanglement for the half-filled
Hubbard model for linear, square and cubic clusters. We note that it
is the itinerancy of electrons that favors the entanglement. Our
results suggest a weak dependence between entanglement and external
spin freedom degrees.

\end{abstract}
 \maketitle

\section{Introduction}

The manipulation of quantum systems in an entangled state that can
be used as a quantum information channel is one of the main
challenges of science today. Information theory, teleportation and
cryptography are just some of the areas that may advance enormously
through the amount of technological applications which can
potentially make use of entanglement. \cite{co1,co2,co3,co4} It is
also of great interest to explore the role played by entanglement
systems in order to understand the basis of quantum mechanics.
\cite{co3,co4} However, even the quantification of entanglement
remains an open question. Current researche is focusing on measures
for precisely quantifying entanglement. \cite{qu1,qu2,qu3} As an
example, entanglement of indistinguishable particles calculated
using different measures has shown that a same quantum state can
have several different characterizations because of the lack of
individual identity of the entangled particles
\cite{in1,in2,in3,grupoufs}.

The use of uncertainty relations has provided an efficient approach
for obtaining one of the most precise experimental measures of
entanglement. \cite{uncert} A quantitative evaluation of the
entangled states can be defined in terms of expectation values of a
convenient witness operator. This operator, called the entanglement
witness, is defined as taking positive values for separable states
and negative ones for entangled states.  Thus, an appropriate
uncertainty relation allows us to choose macroscopic properties
which define an entanglement witness.

A good level of interest has been focused on the special case of
entanglement in macroscopic properties which has been particularly
motivated by experiments that have shown the presence of
entanglement in solid state systems. \cite{cbpf,bene,bruk,bose}.
Wie\'{s}niak {\it et al.} \cite{entang-heis} have recently explored
some aspects of the connection between entanglement and magnetic
susceptibility for an arbitrary Hamiltonian with spin length $s$.

Experimental observations of thermal entanglement in spin systems
using susceptibility measurements have been reported. Souza {\it et
al.} \cite{cbpf} have studied the compound Na$_2$Cu$_5$Si$_4$O$_14$.
They found entanglement confined to the small clusters, with
tripartite entanglement being stronger than bipartite entanglement.
A similar result was obtained by V\'{e}rtesi and Bene \cite{bene} in
the Na$_2$V$_3$O$_7$ system that formed a nanotubular structure of
weakly coupled nine-site rings. Brukner {\it et al.} \cite{bruk} and
Bose and Tribedi \cite{bose} showed entanglement in
antiferromagnetic spin systems.

Since these experimental susceptibility measurements are applicable
to systems with spins localized in sites of the lattice, they are in
accordance with the entanglement witness as defined by Wie\'{s}niak
{\it et al.} \cite{entang-heis} However, this powerful tool is not
adequate for systems with variable local spin lengths, which is an
important feature for conductor materials. The present work
addresses this issue. We will show that new aspects of the
entanglement can appear when we consider systems with variable local
spin lengths. The extension of the entanglement witness across
variable local spins can be related to the itinerant electron models
such as Hubbard and Falikov-Kimball.

In this paper, we investigate the Hubbard model. \cite{hubba} The
purpose was to choose appropriated macroscopic variables in order to
define an entanglement witness adequate for the Hubbard model, or
any other model that can be applied to systems with variable local
spin lengths.

There are some works about entanglement associated to the Hubbard
model.\cite{in1,paolo,gun,larsson,hudak,yao,anfos,revmod} For
example, the entanglement for the Hubbard dimer was investigated by
Dowling {\it et al.} \cite{in1} and Zanardi \cite{paolo}. Gun {\it
et al.} studied the entanglement entropy on the extended Hubbard
model and proposed that the entanglement can be used to identify
quantum phase transitions. \cite{gun} Larsson and Johannesson found
exact expressions for the local entanglement entropy on the
one-dimensional Hubbard model at a quantum phase transition driven
by a change in the magnetic field or chemical potential, related to
the zero-temperature spin and charge susceptibilities.
\cite{larsson} Hudak modeled CeAl$_2$ nanoparticles by the Hubbard
model with negative chemical potential and, using entanglement
entropy he studied the quantum phase transitions present in this
system. \cite{hudak}

Some experimental results have indicated that the entanglement is
restricted to small clusters within the materials. \cite{cbpf,bene}
Exploring this fact, we studied the critical temperature below which
there is thermal entanglement for finite chains and rings, using the
standard direct diagonalization method. \cite{shiba,meuDD} This
approach is very well suited for small sized clusters since it
produces exact results for thermodynamic quantities. Furthermore, it
is also interesting to study the limit of large clusters. In this
case, using the quantum Monte Carlo approach \cite{Hir,meuMC,rai} we
obtained the temperature dependence of the entanglement witness for
linear, square and simple cubic lattices as described by the Hubbard
model. Summarizing, we will show how the cluster length, itinerancy
of the electrons and system dimensions influence the thermal
entanglement on the Hubbard model using direct diagonalization and
quantum Monte Carlo methods.

The organization of this paper is as follows. Entanglement witness
for constant and variable local spin is presented in Sec. II, the
results in Sec. III, and the conclusions in Sec. IV.

\section{Entanglement witness}

The total magnetic susceptibility at null magnetic field
\begin{equation}
\chi = \chi_x + \chi_y + \chi_z = \frac{\left\langle
\vec{M}^{2}\right\rangle - \left\langle \vec{M}
\right\rangle^2}{(\mu_B)^2 N k_B T}
\end{equation}
has been a useful variable to study the witness of thermal
entanglement. Here, $\vec{M}$ is the total magnetization of $N$
spins and $\left\langle...\right\rangle$ is the thermodynamic
average. Considering $s_i$ to be the length of the {\it i}th spin in
the system, the entanglement condition for a thermal state of $N$
spins of same length $s$ ($s_i=s$ for $i=1,...,N$) is given by
\cite{entang-heis}
\begin{equation}
 \chi < \frac{s}{k_B T}.  \label{entcond1}
\end{equation}

The above condition is deduced based on the method of entanglement
detection using the uncertainty relations. \cite{uncert} In summary,
an arbitrary thermal state of spin $s$ has the follow condition
\begin{eqnarray}
 \left\{ \begin{array}{lll}
 \left\langle \vec{S_i}^{2} \right\rangle &=& s(s+1), \\
 \left\langle\vec{S_i}\right\rangle^2 &\le& s^2,  \end{array}
\right.  \label{thermal1}
\end{eqnarray}
where $\vec{S}_i$ is the spin vector of the individual site $i$.
Therefore, if the thermal state is actually a product of $N$ states
of individual spins, the variance of magnetization would be the sum
of variances of individual sites $N k_B T \chi =
\sum_{i=1}^{N}{\left\langle {\vec{S}_i}^{2}\right\rangle -
\left\langle \vec{S}_i \right\rangle^2 } \ge [s(s+1)-s^2] = s$ which
is also valid for the general case of separable states due to the
convexity of the mixture.

However, we notice that the entanglement condition (\ref{entcond1})
fails if the $N$ individual spins have different lengths $s_i \ne
s$. Itinerant systems are an example of this phenomenon because the
$N$ individual sites can have different spin lengths due to the
variety of ways in which they can be filled with particles. It can
also occur in localized systems, since the sites can be filled in
different ways. Thus, the Eq. (\ref{thermal1}) must be generalized
as follows
\begin{eqnarray}
 \left\{ \begin{array}{ccc}
 \sum_{i=1}^{N} \left\langle \vec{S_i}^{2} \right\rangle &=& N\left\langle L_0 \right\rangle, \\
 \left\langle\vec{S_i}\right\rangle^2 &\le& s_{\max}^2,
\end{array} \right.  \label{therma2}
\end{eqnarray}
where $s_{\max}$ is the largest spin length which the individual
sites can take and $L_j \equiv \frac{1}{N} \sum_{i=1}^N
\vec{S_i}\cdot\vec{S}_{i+j}$ is the spin spin correlation function.
Therefore, we can rewrite the condition for entanglement
(\ref{entcond1}) as
\begin{equation}
 \chi < \frac{\left\langle L_0 \right\rangle - s_{\max}^2}{k_B
T}.  \label{entcond2}
\end{equation}

Note that this is also valid for sites with same spin $s_i=s$,
because $\left\langle L_0 \right\rangle = s(s+1)$, $s_{\max} = s$
and consequently the condition above is reduced to Eq.
(\ref{entcond1}).

Particularly, assume an $N$-sites system in which the basis states
are given by
$|n_{1\uparrow},n_{1\downarrow}\rangle\otimes\ldots\otimes|n_{N\uparrow},n_{N\downarrow}\rangle$
where $n_{i\alpha}=$ 0 or 1 (due to the Pauli exclusion principle)
is the number of electrons with $\alpha$-orientation of $S^z$ at the
individual state $i$. Thus, $s_i = 0$ for $|0,0\rangle$ (vacuum
state) or $|1,1\rangle$ (singlet state of two electrons) and $s_i =
1/2$ for $|1,0\rangle$ or $|0,1\rangle$ (single electron states).
Therefore, $s_{\max}=1/2$ and taking into account isotropy $L_j^x =
L_j^y = L_j^z$ (and consequently $\chi^x = \chi^y = \chi^z$), the
generalized condition of thermal entanglement can be expressed as
\begin{equation}
{\cal E} \equiv \chi^z - \frac{\left\langle L_0^z \right\rangle -
1/12}{k_B T} < 0. \label{entfinal}
\end{equation}
Note that if the individual state can only assume single electron
states ($|1,0\rangle$ or $|0,1\rangle$), $s_i=s=1/2$ is fixed,
$\langle L_0^z \rangle = 1/4$ and the condition of entanglement
above reduces to (\ref{entcond1}) as hoped.

The generalization of the entanglement witness for variable local
spins introduces, besides the magnetic susceptibility, the $L_{0}$
as an experimental measurement. Called local moment, the quantity
$L_{0}$ shows the degree of localization of electrons. This
measurement is much less common and more difficult that the magnetic
susceptibility. However, it can be obtained by neutron diffraction
methods. \cite{mezei,keren,ehlers}

\section{Results}

\begin{figure}
\includegraphics[width=0.48\textwidth]{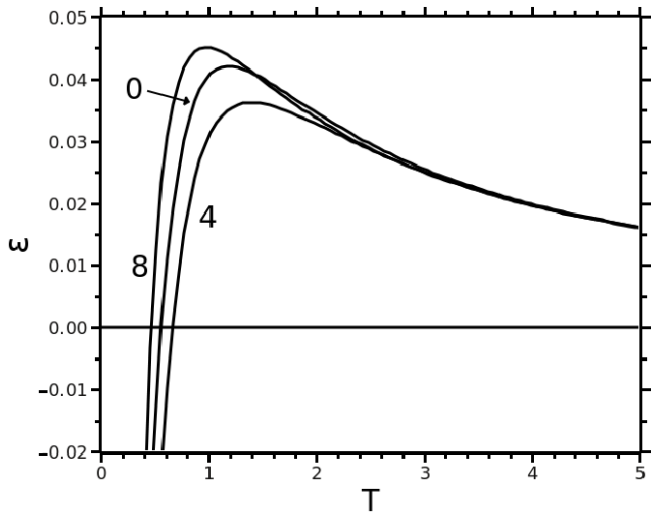}
\caption{Temperature dependence of the witness given by expression
(\ref{entfinal}) for the four site linear chain of the half filled
Hubbard model using the grand canonical ensemble with $U=$ 0, 4 and
8. We adopt units $k_B =1$ and $t=1$.} \label{fig:witness}
\end{figure}

Using the witness (\ref{entfinal}), we investigated the thermal
entanglement for the half filled itinerant electron systems
described by the Hubbard model. The Hamiltonian is
\begin{equation}
{\cal H} = -t \sum_{<ij>\alpha} (c_{i\alpha}^{\dag}c_{j\alpha} + hc)
+ U \sum_{i} n_{i\uparrow}n_{i\downarrow},
\end{equation}
where $c_{i\alpha}^{\dag} (c_{i\alpha})$ are the creation
(annihilation) operators for electrons at site $i$,
$n_{i\alpha}=c_{i\alpha}^{\dag}c_{i\alpha}$, $U$ is the on-site
Coulomb (electron-electron) interaction and $t$ is the nearest
neighbor hopping integral representing the overlap of electron wave
functions.

We have obtained exact results for linear chains and rings with 2, 4
and 6 sites using the numerical method of direct diagonalization of
small clusters over the canonical and the grand canonical
ensembles.\cite{shiba,meuDD} We have observed that the witness
(\ref{entfinal}) for small odd numbers of sites provides no
information about entanglement due to ${\cal E} \gg 0$ for all $T$,
since $\chi^z$ diverges at null temperature. \cite{shiba}

It is illustrated in Fig. \ref{fig:witness} that there is a critical
temperature $T_c$ where ${\cal E}(T_c)=0$ and the system is
entangled for $T < T_c$, because ${\cal E}(T < T_c) < 0$. Therefore,
we can understand $T_c$ as the highest temperature below which the
system is certainly entangled, since there is no certainty about the
entanglement when ${\cal E} (T \ge T_c) \ge 0$. \cite{uncert}
\begin{figure}
\includegraphics[width=0.48\textwidth]{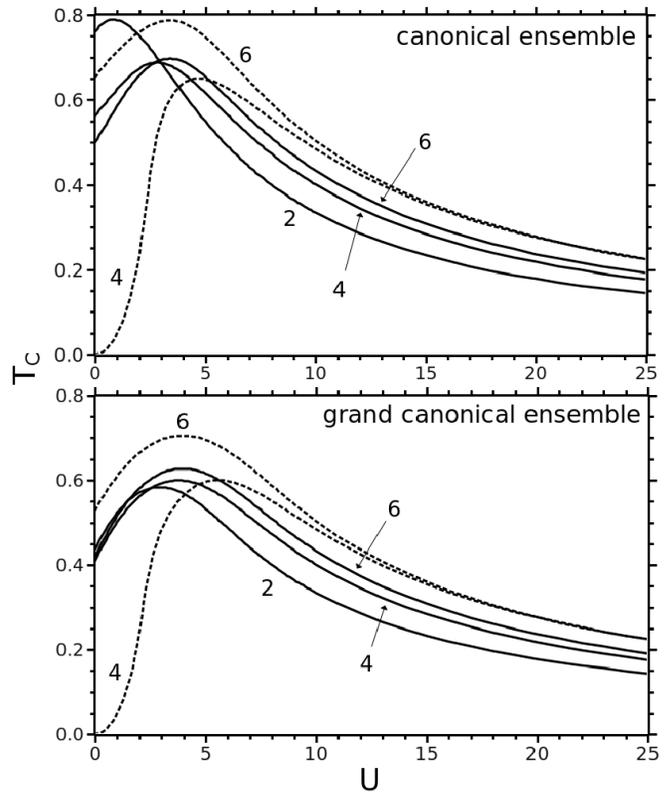}
\caption{Coulombian interaction dependence of the critical
temperature for the finite one-dimensional half filled Hubbard
model. We adopt units $k_B =1$ and $t=1$. The solid and dashed lines
are related to chains and rings, respectively. Each curve is labeled
by its number of sites.} \label{fig:tcxu-dd-1d}
\end{figure}

Fig. \ref{fig:tcxu-dd-1d} exhibits $T_c$ versus $U$ for 1D systems.
A comparison between results for different ensembles shows a good
agreement at large Coulomb interaction, but not at small ones. We
have found a rich dependence on $U$ and $N$. With fixed $N$ for
small $U/t$, we see that the value of $T_c$ increases as the value
of $U/t$ is increased and tends to a maximum value. All curves have
presented a value of interaction $U$ which produces the maximum
$T_c$ (global maximum of $T_c$ vs. $U$). For the strong Coulombian
interaction $U \gg t$, we notice that all curves present a
hyperbolic behavior $T_c \propto U^{-1}$. We will define the
parameter $\eta(N,U) \equiv A_N k_B T_c / (4t^2/U)$, where $A_N$ is
a function of $N$. $\eta(N,U)$ is convenient to compare our results
with the Heisenberg model ones, considering that in the asymptotic
regime $U \gg t$ there is an equivalence between the half filled
Hubbard and the Heisenberg models with exchange interaction
$J=4t^2/U$. \cite{emery}

For the grand canonical ensemble, as the size of an even sites
system increases, the values of the maximum global $U^{\max}$ and
$T_c^{\max}$ also increase. Although the canonical ensemble has a
similar increasing relation between $U^{\max}$ and $N$, there is no
monotonic behavior of $T_c^{\max}$ versus $N$. We performed a
numerical extrapolation using the grand canonical ensemble for
linear chains with 2, 4 and 6 sites. Our extrapolation analysis
predicts $k_BT_c^{\max} = 0.712t$ at $U^{\max}= 4.1t$ in the
thermodynamic limit. We also obtained $\eta(\infty,\infty) \cong
1.568 \pm 0.003$ which is very close to the exact value
$\eta(\infty,\infty) = k_B T_c / J = 1.6$ for $\frac{1}{2}$-$s$
Heisenberg model. \cite{entang-heis}

\begin{figure}[h]
\includegraphics[width=0.48\textwidth]{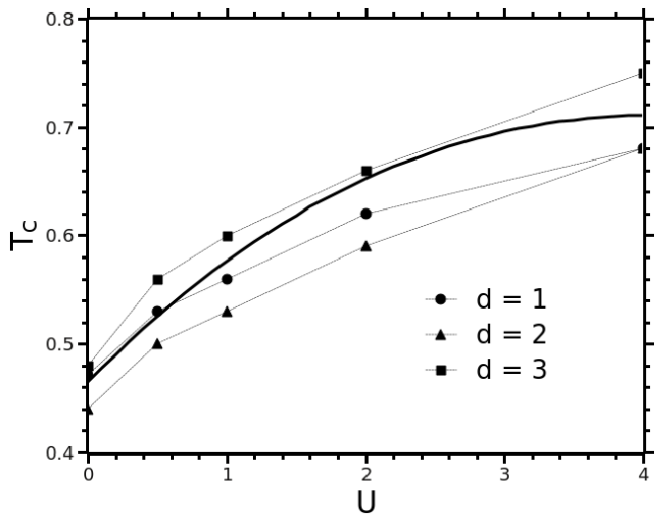}
\caption{Coulombian interaction dependence of the critical
temperature for the one, two and three dimensional half filled
Hubbard model. We adopt units $k_B =1$ and $t=1$. The scatter is
related to the quantum Monte Carlo method and the solid line is an
extrapolation of the thermodynamic limit for a linear chain.}
\label{mc}
\end{figure}

The direct diagonalization approach is very suitable for small sized
clusters, but becomes inefficient when the system has its size
increased. On the other hand, the quantum Monte Carlo (QMC) method
\cite{Hir,meuMC} is an efficient approach to study large systems.
Using it we have studied the entanglement witness for linear, square
and cubic lattices.

The QMC method treats the exponentials of the grand partition
function with the Suzuki-Trotter decomposition scheme. Using a
discrete Hubbard-Stratonovich transformation it converts the
electron-electron interaction into one of free electrons interacting
with a time-dependent Ising field. With it we compute the relative
weights of the Ising field configurations. The algorithm follows the
lines of those for classical systems, except for the Boltzmann
weight that is expressed as a sum over Ising spins of a product of
determinants. For the Hubbard model at half filling, the product of
determinants is always positive. We have used the imaginary time
discretization of the QMC $\Delta \tau =0.125$. \cite{rai}


Fig. \ref{mc} shows the QMC results for a 64-site ring, and for
100-site square and cubic lattices. Our $T_c$ cannot be estimated
accurately for large $U/t$ because the QMC becomes unstable at low
temperatures and with strong Coulombian interaction. \cite{rai}

Furthermore, we include in Fig. \ref{mc} the thermodynamic limit
extrapolation obtained through the linear chain results from the
direct diagonalization for small clusters. Note that the
extrapolation is consistent with the simulation. Notice that for
different lattices, the results for $T_c$ are similar revealing no
new behavior.

\section{CONCLUSION}

Wesniak \textit{et al.} \cite{entang-heis} have suggested that
magnetic susceptibility can be a macroscopic (thermodynamical) spin
entanglement witness without complete knowledge of the specific
model (Hamiltonian) of the solid. However, we observed here that its
applicability is restricted to the insulator materials because local
features of the spin length affect the deviation of the witness. We
have developed a generalization that allows us to get information
about system entanglement with variable local spin lengths such as
found in itinerant electron systems. Moreover, our witness is also
valid for fixed local spin lengths and consequently, it can be used
experimentally in conductor or insulator materials.

As an application, we studied thermal entanglement for the one, two
and three dimensional half filled Hubbard model.  We obtained the
critical temperature $T_c$ below which the system is certainly
entangled. We have shown that there is a Coulombian repulsion that
presents a global $T_c$ maximum. This feature is relevant for
quantum information science, since it reveals the optimal Coulombian
repulsion referent to the highest temperature where the system is
definitely entangled. In addition, the decrease of $T_c$ for $t \ll
U$ indicates that the itinerancy of electrons favors the
entanglement. Furthermore, at the asymptotic regime $U \gg t$ we
show, through a numerical extrapolation to the thermodynamic limit,
that $T_c$ is in accordance with the exact result for the
$\frac{1}{2}$-$s$ Heisenberg model. A recent study has shown that
higher spin length increases the $T_c$.  \cite{entang-heis} Since
higher spin length means higher internal degrees of freedom the
above result shows a strong favoring of entanglement according to
the increases in the internal degrees of freedom. In this work, from
the results of $T_c$ for linear, square and cubic lattices, we
notice that an increase in the external spin degrees of freedom
produces similar results. These results suggest that the dependence
between entanglement and internal spin degrees of freedom is far
stronger than between entanglement and external spin degrees of
freedom.

\section*{ACKNOWLEDGMENTS}
This work was supported by CNPq (Brazil).


\begin{thebibliography}{25}
\bibitem{co1} C. E. Mora and H. J. Briegel, Phys. Rev. Lett. {\bf 95}, 200503 (2005).
\bibitem{co2} Y. Shimoni, D. Shapira, and O. Biham, Phys. Rev. A {\bf 72}, 062308 (2005).
\bibitem{co3} G. Alber, T. Beth, P. Horodecki, R. Horodecki, M.
R\"ottler, H. Weinfurter, R. Werner, A. Zeilinger, {\it Quantum
Information} (Springer Tracts in Modern Physics, Vol. 173,  Berlin,
2001).
\bibitem{co4} M. A. Nielsen and I. L. Chuang, {\it Quantum Computation and Quantum Information}
(Cambridge University Press, Cambridge, 2000).
\bibitem{qu1} C. S. Yu and H. S. Song, Phys. Rev. A {\bf 73}, 022325 (2006).
\bibitem{qu2} D. Larsson and H. Johannesson, Phys. Rev. A {\bf 73}, 042320 (2006).
\bibitem{qu3} K. Chen, S. Albeverio, and S.-M. Fei, Phys. Rev. Lett. {\bf 95}, 210501 (2005).
\bibitem{in1} M. R. Dowling, A. C. Doherty, and H. M. Wiseman, Phys. Rev. A {\bf 73}, 052323 (2006).
\bibitem{in2} H. M. Wiseman and John A. Vaccaro, Phys. Rev. Lett. {\bf 91}, 097902 (2003).
\bibitem{in3} G. C. Ghirardi, L. Marinatto, Phys. Rev. A {\bf 70}, 012109 (2004).
\bibitem{grupoufs} V. C. G. Oliveira, H. A. B. Santos, L. A. M. Torres, and A. M. C.
Souza, Int. J. Quantum Inf. {\bf 6}, 379 (2008).
\bibitem{uncert} H. F. Hofmann and S. Takeuchi, Phys. Rev. A {\bf 68}, 032103 (2003).
\bibitem{cbpf} A. M. Souza, M. S. Reis, D. O. Soares-Pinto, I. S. Oliveira, and R. S. Sarthour, Phys. Rev. B {\bf 77}, 104402 (2008).
\bibitem{bene} T. V\'{e}rtesi and E. Bene, Phys. Rev. B {\bf 73}, 134404 (2006).
\bibitem{bruk} \v{C}. Brukner, V. Vedral, and A. Zeilinger, Phys. Rev. A {\bf 73}, 012110 (2006).
\bibitem{bose} I. Bose and A. Tribedi, Phys. Rev. A {\bf 72}, 022314 (2005).
\bibitem{entang-heis} M. Wie\'{s}niak, V. Vedral and \v{C}. Brukner, New J. Phys. {\bf 7}, 258 (2005).
\bibitem{hubba} J. Hubbard, Proc. R. Soc. (London) A {\bf 276}, 238 (1963).
\bibitem{paolo} P. Zanardi, Phys. Rev. A {\bf 65}, 042101 (2002).
\bibitem{gun} S. J. Gu, S. S. Deng, Y. Q. Li, and H. Q. Lin, Phys. Rev. Lett. {\bf 93}, 086402 (2004).
\bibitem{larsson} D. Larsson and H. Johannesson, Phys. Rev. Lett. {\bf 95}, 196406 (2005); Phys. Rev. Lett. {\bf 96}, 169906(E)
(2006).
\bibitem{hudak} O. Hudak, Phys. Lett. A. {\bf 373}, 359 (2008).
\bibitem{yao} K.L. Yao, Y.C. Li, X.Z. Sun, Q.M. Liu, Y. Qin, H.H. Fu and G.Y. Gao, Phys. Lett. A. {\bf 346}, 209 (2005).
\bibitem{anfos} A. Anfossi, P. Giorda, and A. Montorsi, Phys. Rev. B {\bf 75}, 165106
(2007).
\bibitem{revmod} L. Amico, R. Fazio, A. Osterloh, and V. Vedral, Rev. Mod. Phys.
{\bf 80}, 517 (2008).
\bibitem{shiba} H. Shiba and P. A. Pincus, Phys. Rev. B {\bf 5}, 1966 (1972).
\bibitem{meuDD} C. A. Macedo and A. M. C. de Souza, Phys. Rev. B {\bf 65}, 153109 (2002).
\bibitem{Hir} J. E. Hirsch, Phys. Rev. B {\bf 31}, 4403 (1985).
\bibitem{meuMC} C. A. Macedo and A. M. C. Souza, Physica B {\bf 354}, 290 (2004);
A. M. C. Souza and C. A. Macedo, J. Magn. Magn. Mater. {\bf 226},
2026 (2001).
\bibitem{rai} R. R. dos Santos, Braz. J. Phys. {\bf 33}, 36 (2003).
\bibitem{mezei} F. Mezei and A. P. Murani, J. Magn. Magn. Mater. {\bf 14}, 211 (1979). 
\bibitem{keren} A. Keren, P. Mendels, I. A. Campbell, and J. Lord, Phys. Rev. Lett. {\bf 77}, 1386 (1996). 
\bibitem{ehlers} G. Ehlers, J. Phys.: Condens. Matter {\bf 18}, R231 (2006). 
\bibitem{emery} V. J. Emery, Phys. Rev. B {\bf 14}, 2989 (1976).
\end{thebibliography}
\end{document}